\documentclass[12pt]{article}
\usepackage{amsmath,amsfonts,amssymb,latexsym,graphicx,caption}
\DeclareGraphicsExtensions{.pdf,.jpg,.PNG,.JPG}

\numberwithin{equation}{section}
\def\be{\begin{equation}}
\def\ee{\end{equation}}
\def\bq{\begin{eqnarray}}
\def\eq{\end{eqnarray}}
\def\beq{\begin{eqnarray}}
\def\eeq{\end{eqnarray}}

\def\pa{\partial}

\begin{document}

\title{\Large{\textsc{Regular braneworlds with bulk fluids}}}
\author{{\large\textsc{Ignatios Antoniadis$^{1,2}$\thanks{\texttt{antoniad@lpthe.jussieu.fr}}
Spiros Cotsakis$^{3,4}$\thanks{\texttt{skot@aegean.gr}}
Ifigeneia Klaoudatou$^{4}$\thanks{\texttt{iklaoud@aegean.gr}}}} \\
$^1$Laboratoire de Physique Th\'{e}orique et Hautes Energies - LPTHE\\
Sorbonne Universit\'{e}, CNRS 4 Place Jussieu, 75005 Paris, France\\
$^2$ Institute for Theoretical Physics, KU Leuven\\ Celestijnenlaan 200D, B-3001 Leuven, Belgium\\
$^{3}$Institute of Gravitation and Cosmology, RUDN University\\
ul. Miklukho-Maklaya 6, Moscow 117198, Russia\\
$^{4}$Research Laboratory of Geometry,  Dynamical Systems  and Cosmology\\
University of the Aegean, Karlovassi 83200, Samos, Greece}
\maketitle
\newpage
\begin{abstract}
\noindent
We review studies on the singularity structure and
asymptotic analysis of a 3-brane (flat or curved) embedded in a
five-dimensional bulk filled with a `perfect fluid’ with
an equation of state $p=\gamma\rho$, where $p$ is the ‘pressure’ and $\rho$ is the `density’
of the fluid, depending on the 5th space coordinate.
Regular solutions satisfying positive energy conditions in the bulk exist only in the cases of a flat brane
for $\gamma=-1$ or of AdS branes for $\gamma\in [-1,-1/2)$. More cases can be found by gluing two
regular brunches of solutions at the position of the brane. However, only a flat brane
for $\gamma=-1$ leads to finite Planck mass on the brane and thus localises gravity.
In a more recent work, we showed that a way to rectify the previous findings and obtain
a solution for a flat brane and a range of $\gamma$, that is both free from finite-distance singularities
and compatible with the physical conditions of energy and finiteness of
four-dimensional Planck mass, is by introducing a bulk fluid component that satisfies a non-linear
equation of state of the form $p=\gamma\rho^{\lambda}$ with $\gamma<0$ and $\lambda>1$.
\end{abstract}
\newpage
\tableofcontents
\newpage
\section{Introduction}
In this paper, we review the dynamical evolution and physical properties of
a class of higher-dimensional models analysed in \cite{ack1, ack2, ack3, ack4, ack5, ack6, ack7}.
Our motivation for studying higher-dimensional models stems from the fact
that they propose alternative approaches towards understanding and
hopefully improving our view on challenging issues in cosmology and particle physics.
In this review specifically, we focus on a class of brane-world models that offers interesting
implications on the cosmological constant problem (hereafter abbreviated cc-problem).

The cc-problem arises from the disagreement between predictions from
quantum field theories and observations regarding the value of the cosmological constant. In particular, the theoretical 
quantum corrections to the cosmological constant are naturally some 120 orders of
magnitude higher than its observed value. Based on theory, the huge value
of the cosmological constant, would automatically imply a huge value of the
vacuum energy which 
would in turn give rise to a highly curved universe, a prediction that is not
compatible with observations. To resolve the discrepancy
between theory and observations, the bare value of the
cosmological constant has to be unreasonably fine tuned.

A viable approach to this puzzling problem, is to re-examine it in the context of
higher-dimensional models. A first study towards this way of resolving the cc-problem,
was explored in \cite{rubakov}. The main idea was that in the framework of higher
dimensions (4+2, where 2 are extra compactified a la Kaluza-Klein spatial dimensions),
the cosmological constant could only curve the extra dimensions, leaving the
four-dimensional observed universe (almost) flat. Following this scenario,
a class of solutions with arbitrary (including zero) values of the observable
cosmological constant was found in \cite{rubakov}. However, the question of why the
solutions with a vanishing cosmological constant should be singled out was kept unanswered,
while various factors that could favour these solutions were discussed, such as
appropriate quantum corrections and/or additional interactions, or, even
stability considerations.

Later on, the idea to use extra dimensions to attack the cc-problem,
was revisited in \cite{nima, silverstein, forste1, forste2, forste3, csaki,gubser}. The
setup of these models is based on the idea that the
observed universe is modeled by a four-dimensional hypersurface situated at a fixed position of an extra spatial
dimension. The whole spacetime is five dimensional: there are four dimensions
of space, out of which only three are spanned by the hypersurface
and one dimension of time. Such a hypersurface is called 3-brane
while the full higher-dimensional spacetime is called the bulk. Branes play an important role in string theory as they are the designated
locations where open strings end and at the same time they are essential in
proving duality between different versions of the theory. Additional
interest on brane-worlds was sparked from the realization of a novel proposal
towards resolving the hierarchy problem \cite{ant1,add,aadd, rs1, rs2,ant2}
within their context.

In the brane-world scenario, the brane behaves like a thin surface layer
that is embedded in the bulk and it is associated with a surface-energy momentum
tensor that acts as a delta-source of matter. The surface-energy momentum tensor
creates a jump in the extrinsic curvature and describes all Standard Model fields.
Gravity and some non-standard fields on the other hand, experience also
the fifth spatial dimension and interact with the fields on the brane, through
a coupling function (tension) that depends on the setup of each model. In particular, the
specific class of models that we review here, have a warped geometry with a
line-element of the form
\be
ds^{2}=a^{2}(Y)g_{4}+dY^{2}
\ee
where $Y$ is the extra dimension, $g_{4}$ is the 4D metric and $a(Y)>0$ is the
warp factor that describes deformations with respect to the extra dimension. The
embedding of the brane in the bulk, introduces a $Y\rightarrow -Y$ mirror
symmetry and a set of Israel junction conditions \cite{israel1, israel2} that relate the tension of
the brane with the value of the warp factor at the position of the brane (chosen at the origin of $Y$).
This latter relation plays a key role as we discuss below, since it implicitly
connects the tension of the brane with the value of the 4D scalar curvature.

Refs. \cite{nima} and \cite{silverstein} use higher-dimensional models in an
effort to ameliorate the cc-problem. Their proposal is to explore the
possibility of a self-tuning mechanism. To understand better this mechanism,
we have to keep in mind that the tension of the brane receives quartically divergent
quantum corrections from vacuum energy and through the Israel conditions
which, as mentioned above, are essentially boundary conditions describing
the embedding of the brane in the bulk, transfers these corrections to the
curvature of the brane.
However, if it is possible to find flat-brane solutions irrespectively from the
fluctuations of the brane tension, then we could end-up with a universe that is
self-tuned to a vanishing cosmological constant on the brane.

In \cite{nima}, the bulk matter is modeled by a scalar field that is minimally coupled to gravity
and conformally coupled to the fields on the brane. The conformal coupling is
carefully chosen to allow only for a flat brane in support to the self-tuning
mechanism. In \cite{silverstein} on the other hand, the bulk matter can also
contain a 5D cosmological constant and a variety of forms of tension.

A common feature of both models of \cite{nima} and \cite{silverstein},
is the emergence of singularities that arise within finite distance from the
position of the brane. While investigating the puzzling nature of the
finite-distance singularities, it was argued initially in \cite{nima} that these
singularities, on one hand, can be viewed to act like a reservoir through which
all the vacuum energy is emptied, while on the other hand,
can serve at successfully compactifying the extra dimension. Thus, it was
implied that the existence of such singularities serves in achieving both
4D gravity macroscopically and a vanishing cosmological constant. The
details of a mechanism that could lead to a smooth transition to 4D dynamics
was not further explored.

Later works \cite{forste1} however, showed that the flat-brane
solutions of \cite{silverstein} containing finite-distance singularities fail to meet a consistency
condition that would ensure that the field equations are globally satisfied.
This result was extended in \cite{forste2} by considering a variety
of vacuum configurations. As it was pointed out in \cite{forste1, forste2},
to obtain consistency, the singularities have to be resolved. One way to
achieve this, is by introducing extra branes at the positions
of the singularities. Unfortunately, embedding more branes entails defining
new boundary conditions which in turn introduces a type of fine tuning
in the model. An alternative way to rectify the singularities, is by exploiting the
mirror symmetry introduced by the embedding of the brane in the bulk
and choosing the parameters of the model appropriately to construct
a regular matching solution, by cutting and matching the part
of the bulk that does not contain singularities. Again, this leads to more issues, since the matching solution gives
an infinite Planck mass, thus failing to localize gravity on the brane.

A way to overcome this was explored in \cite{forste3}, with the use of a bulk
scalar field with an unorthodox Lagrangian. It was shown that by choosing
the range of parameters appropriately and using the matching mechanism
mentioned above, it is possible to avoid finite-distance singularities and
construct a flat-brane solution that is both regular and at the same time
suitable for localizing gravity on the brane. However, the derived solution
faces stability issues.

In \cite{csaki}, the possibility of finding general forms of bulk potentials
leading to self-tuned solutions without a finite-distance singularity and being
valid for a range of brane tensions compatible with localized gravity
on the brane, was explored. It was found that singularities in self-tuned
solutions are generic for localizing gravity on the brane. Resolving the
singularities essentially re-introduces fine-tuning in accordance with
\cite{forste1, forste2}. However, it was
suggested that additional fields may help revive the self-tuning mechanism.

In \cite{ack1, ack2, ack3, ack4, ack5, ack6, ack7}, we generalized the work of \cite{nima} by modeling
the bulk matter with a variety of components such as an analog of a perfect
fluid $p=\gamma\rho$, where the `pressure' $p$ and the `density' $\rho$ are
functions of the extra dimension only, an interacting mixture, or more recently, a non-linear fluid with equation of state
$p=\gamma\rho^\lambda$. In most of these cases our models contained also a curved
brane. We find that the non-linear equation of state
$p=\gamma\rho^\lambda$, is the most appropriate type of bulk matter
for generating regular solutions with fundamental physical properties.
Such an equation of state has been studied previously in cosmology for its
role in avoiding big-rip singularities during late time asymptotics
\cite{srivastava, diaz, nojiri}, obtaining inflationary models with special
properties \cite{ba90}, unifying models of dark energy and dark matter
\cite{ka, sen}, but also in the analysis of singularities \cite{ck1, ck2, not}.

The mathematical tools that we used for performing our analysis
in \cite{ack1, ack2, ack3, ack4, ack5, ack6, ack7},
include the method of asymptotic splittings \cite{dom} that detects all possible
asymptotic behaviors of solutions around a singularity, combined with a
method for tracing envelopes, cf. \cite{ack4}, Section 2,  which are essentially solutions with
a smaller number of arbitrary constants but which, nonetheless, play a crucial role in determining
the general behavior of solutions. Another
tool is the analysis of asymptotic behaviors of Gaussian hypergeometric
functions that arise in solutions of curved branes, or, even of flat branes for a
non-linear bulk fluid. In this paper, we overview the basic mathematical details and physical
implications of the body of work of \cite{ack1, ack2, ack3, ack4, ack5, ack6, ack7}.

The structure of this paper is as follows.
In Section 2, we give the setup and field equations for our brane-worlds.
In Section 3, we study the weak, strong and null energy conditions for the
bulk fluid. Next, in Section 4, we overview the case of a linear fluid for a flat, or,
curved brane and analyse the corresponding energy conditions, as well as,
the possibility of localizing gravity on the brane. Then, in Section 5, we study
from the same perspective,  a non-linear fluid,
first, for $\lambda=3/2$ and then for general $\lambda$. Finally, in Section 6,
we present our conclusions and discuss open questions.
\section{Setup and field equations}
We study a class of brane-world models that consist of a flat, or, a curved 3-brane
embedded in a five-dimensional bulk. The bulk metric is given by
\be
\label{warpmetric}
g_{5}=a^{2}(Y)g_{4}+dY^{2},
\ee
where $g_{4}$ is the four-dimensional flat, de Sitter or anti de Sitter metric,
{\it i.e.},
\be
\label{branemetrics}
g_{4}=-dt^{2}+f^{2}_{k}g_{3},
\ee
with
\be
\label{g_3}
g_{3}=dr^{2}+h^{2}_{k}g_{2},
\ee
and
\be
\label{g_2}
g_{2}=d\theta^{2}+\sin^{2}\theta d\varphi^{2},
\ee
with $f_{k}=1,\cosh (H t)/H,\cos (H t)/H$ ($H^{-1}$ is the de Sitter (or AdS)
curvature radius), $h_{k}=r,\sin r,\sinh r$, respectively and
$a(Y)$ is the warp factor ($a(Y)>0$) which we simply denote by $a$.

In our notation, capital Latin indices take the values $A,B,\dots=1,2,3,4,5$,
while lowercase Greek indices are taken to range as $\alpha,\beta,\ldots =1,2,3,4$, with
$t$ being the timelike coordinate, $(r,\theta,\phi,Y)$ the remaining spacelike ones
and the 5th coordinate corresponding to $Y$. The 5-dimensional Riemann
tensor is defined by the formula,
\be
R^{A}_{\,\,\,BCD}=\partial_{C}\Gamma^{A}_{\,\,\,BD}-\partial_{D}\Gamma^{A}_{\,\,\,BC}+\Gamma^{M}_{BD}\Gamma^{A}_{MC}-\Gamma^{M}_{BC}\Gamma^{A}_{MD}
\ee
the Ricci tensor is the contraction,
\be
R_{AB}=R^{C}_{\,\,\,ACB},
\ee
and the five-dimensional Einstein equations on the bulk space are given by,
\be
G_{AB}=R_{AB}-\frac{1}{2}g_{AB}R=\kappa^{2}_{5}T_{AB}.
\ee

We assume that the bulk is filled with a fluid analogue with energy-momentum tensor
of the form
\be
\label{T old}
T_{AB}=(\rho+p)u_{A}u_{B}-pg_{AB},
\ee
where the `pressure' $p$ the `density' $\rho$ are functions only of the fifth
dimension, $Y$, and the velocity vector field is $u_{A}=(0,0,0,0,1)$, that is
$u_{A}=\pa/\pa Y$, parallel to the $Y$-dimension.
The five-dimensional Einstein equations can then be written as
\bq
\label{syst2i general}
\frac{a'^{2}}{a^{2}}&=&\frac{\kappa_{5}^{2}}{6}\rho+\frac{k H^{2}}{a^{2}},\\
\label{syst2ii general}
\frac{a''}{a}&=&-\frac{\kappa_{5}^{2}}{6}(\rho+2p),
\eq
where $k=\pm 1$, and the prime $(\,')$ denotes differentiation with
respect to $Y$. On the other hand, the equation of conservation,
$$
\nabla_{B}T^{AB}=0,
$$
gives
\be
\label{syst2iii general}
\rho'+4\frac{a'}{a}(\rho+p)=0.
\ee

We assume that the density and pressure of the fluid are related according to
the general equation of state
\be
\label{eos}
p=\gamma\rho^{\lambda},
\ee
where $\gamma$ and $\lambda$ are constants. Inputting (\ref{eos}) in
(\ref{syst2i general})-(\ref{syst2iii general}), we find
\bq
\label{syst2i}
\frac{a'^{2}}{a^{2}}&=&\frac{\kappa_{5}^{2}}{6}\rho+\frac{k H^{2}}{a^{2}},\\
\label{syst2ii}
\frac{a''}{a}&=&-\frac{\kappa_{5}^{2}}{6}(\rho+2\gamma\rho^{\lambda}),
\eq
\be
\label{syst2iii}
\rho'+4\frac{a'}{a}(\rho+\gamma\rho^{\lambda})=0.
\ee

Before solving the system of Eqs.~(\ref{syst2i})-(\ref{syst2iii})
and studying its asymptotic behaviors, we find it useful to outline below, the
possible types of singularity that we will encounter in the next Sections.
Denoting with $Y_{s}$, the finite value of $Y$ labeling the position of the
singularity, we say that a finite-distance singularity is a
\begin{itemize}
\item {\em collapse} singularity, if $a\rightarrow 0^{+}$, as $Y\rightarrow Y_{s}$,

\item {\em big-rip} singularity, if $a\rightarrow\infty$, as $Y\rightarrow Y_{s}$.
\end{itemize}
Depending on 
the values of $k$, $\gamma$ and $\lambda$, the above behaviors may be accompanied by a divergence in
the density, or, even in the pressure of the fluid.
We emphasize that, these singularities are not related to geodesic
incompleteness as in standard cosmology, but rather on a pathological
behavior of the warp factor. In the absence of finite-distance singularities, we
call the solutions {\em regular} and include in this category the behaviors of the warp factor given above, provided
that these occur {\em only} at infinite distance, {\em i.e} $Y\rightarrow\pm \infty$.

Our study will be completed once we find a solution that has the following
fundamental properties:
\begin{itemize}
\item it is regular (no finite-distance singularities)
\item it satisfies physical conditions, such as energy conditions
\item it leads to a finite Planck mass, thus, it localizes gravity on the brane.
\end{itemize}

Since the behaviors of solutions depend strongly on the curvature of the brane, as well as on the linearity/non-linearity
of the equation of state, we present the various possibilities in separate
Sections. Also, in the next Section, we explain briefly a way to formulate the weak, strong and null energy
condition for our type of bulk matter.
Detailed proofs of our results can be found in \cite{ack4, ack6}.
\section{Energy conditions}
The weak, strong and null energy condition are physical requirements that
we wish our solutions to fulfill. They can help us single out those solutions of
Eqs.~ (\ref{syst2i})-(\ref{syst2iii}) as more plausible from a physics
point of view. Classically, it is convenient to translate the energy conditions
to restrictions imposed on $p$ and $\rho$.

To work out the energy conditions for our type of fluid, we start by
noting that in the formulation of the field equations, both the metric
given by (\ref{warpmetric}) and the bulk fluid described by
(\ref{T old}) and (\ref{eos}), appear as static with respect to the time
coordinate $t$, because the evolution is taken with respect to the fifth
spatial coordinate $Y$. Using
this fact, we can reinterpret our fluid analogue as a real anisotropic fluid
having the following energy momentum tensor,
\be
\label{T new}
T_{AB}= (\rho^{0}+p^{0})u_{A}^{0}u_{B}^{0}-p^{0}g_{\alpha\beta}\delta_{A}^{\alpha}\delta_{B}^{\beta}-
p_{Y}g_{55}\delta_{A}^{5}\delta_{B}^{5},
\ee
where $u_{A}^{0}=(a(Y),0,0,0,0)$, $A,B=1,2,3,4,5$ and $\alpha,\beta=1,2,3,4$.
When we combine (\ref{T old}) with (\ref{T new}), we get the following set of
relations,
\bq
\label{p y to rho}
p_{Y}&=&-\rho\\
\label{rho new}
\rho^{0}&=&-p\\
\label{p new}
p^{0}&=&p.
\eq
Note that, the last two relations imply that
\be
\label{p new to rho new}
p^{0}=-\rho^{0},
\ee
which means that this type of matter satisfies a cosmological constant-like equation of state.
Substituting (\ref{p y to rho})-(\ref{p new to rho new})
in (\ref{T new}), we find that
\be
T_{AB}= -p g_{\alpha\beta}\delta_{A}^{\alpha}\delta_{B}^{\beta}+
\rho g_{55}\delta_{A}^{5}\delta_{B}^{5}.
\ee

At this point, we can start to formulate the energy conditions. First, we
study the weak energy condition according to which, every future-directed
timelike vector $v^{A}$ should satisfy
\be
T_{AB}v^{A}v^{B}\geq 0.
\ee
This condition implies that the energy density should be non negative for all forms of
physical matter \cite{wald}. Here we find 
that it translates to
\be
\label{wec_p}
p\geq 0, 
\ee
and
\be
\label{wec}
p+\rho\geq 0.
\ee
%
Second, we work out the strong energy condition which states that
\be
\left(T_{AB}-\dfrac{1}{3}T g_{AB}\right)v^{A}v^{B}\geq 0,
\ee
for every future-directed unit timelike vector $v^{A}$. In our case,
\be
\label{sec_1}
-p+\rho\geq 0,
\ee
and
\be
\label{sec_2}
p+\rho\geq 0.
\ee

Finally, we study the null energy condition according to which, every future-directed
null vector $k^{A}$ should satisfy \cite{poisson}
\be
T_{AB}k^{A}k^{B}\geq 0.
\ee
Here we find that it translates to
\be
\label{nec}
p+\rho\geq 0.
\ee
In later Sections we are going to express these conditions with respect to
the values of the parameters $\gamma$ and $\lambda$ of the equation
of state, as this will enable us to automatically recognize those solutions
that are compatible with the energy conditions.
\section{Linear fluid}
In this section, we review the behaviors of solutions for a linear fluid,
which can be also viewed as an analogue of a `perfect' fluid.

Inputting $\lambda=1$ in the field equations (\ref{syst2i})-(\ref{syst2ii}),
we find
\bq
\label{syst2i-linear}
\frac{a'^{2}}{a^{2}}&=&\frac{\kappa_{5}^{2}}{6}\rho+\frac{k H^{2}}{a^{2}},\\
\label{syst2ii-linear}
\frac{a''}{a}&=&-\frac{\kappa_{5}^{2}}{6}(1+2\gamma)\rho,
\eq
while (\ref{syst2iii}) gives
\be
\label{syst2iii-linear}
\rho'+4(1+\gamma)\frac{a'}{a}\rho=0.
\ee


Naturally, the forms of solutions of Eqs.~(\ref{syst2i-linear})-(\ref{syst2iii-linear}),
depend strongly on the values of $k$ and $\gamma$.
We classify the types of solutions and examine
each class, separately, in the following subsections.
\subsection{Flat brane}
To study the case of a flat brane, we first substitute $k=0$ in
(\ref{syst2i-linear}) and find
\be
\label{syst2i-linear-flat}
\frac{a'^{2}}{a^{2}}=\frac{\kappa_{5}^{2}}{6}\rho.
\ee
Next, we integrate (\ref{syst2iii-linear}) to obtain the relation between
$\rho$ and $a$ which reads
\be
\label{rho to a-linear}
\rho=c_{1}a^{-4(\gamma+1)},
\ee
with $c_{1}$ an arbitrary constant.

We can now substitute (\ref{rho to a-linear}) in (\ref{syst2i-linear-flat}) and
integrate to derive the form of the warp factor, $a$,
\be
\label{solution a-linear-flat}
a=\left(2(\gamma+1)\left(\pm\sqrt{\dfrac{2Ac_{1}}{3}}Y+c_{2}\right)\right)^{1/(2(\gamma+1))}, \quad \gamma\neq -1,
\ee
where $A=\kappa_{5}^{2}/4$.
Finally, we input (\ref{solution a-linear-flat}) in (\ref{rho to a-linear}) to
find $\rho$:
\be
\label{solution rho-linear-flat}
\rho=c_{1}\left(2(\gamma+1)\left(\pm\sqrt{\dfrac{2Ac_{1}}{3}}Y+c_{2}\right)\right)^{-2}.
\ee
Substitution of our solution for $k=0$ of $a$ and $\rho$ in (\ref{syst2ii-linear})
shows that the latter equation is satisfied.

Our solution (\ref{solution a-linear-flat}) and (\ref{solution rho-linear-flat}) holds
for all values of $\gamma$ except from $\gamma=-1$. The case of $\gamma=-1$
is a special one and is studied separately in a following subsection.
For all other values of $\gamma$, we see that for a flat brane
and a linear equation of state, there is always a finite-distance singularity
located at $Y_{s}=\mp c_{2}\sqrt{3/(2A c_{1})}$. The nature of the singularity
depends on whether $\gamma$ is less, or, greater than $-1$, and can be classified
into a collapse type of singularity with
\be
a\rightarrow 0, \quad \rho\rightarrow\infty,\quad Y\rightarrow \mp c_{2}\sqrt{3/(2A c_{1})}, \quad \gamma>-1,
\ee
or, big-rip type with
\be
a\rightarrow \infty,\quad \rho\rightarrow\infty, \quad Y\rightarrow \mp c_{2}\sqrt{3/(2A c_{1})},\quad \gamma<-1.
\ee

Combining this outcome with similar results found in \cite{nima} for a massless
scalar field, which can be also viewed as a fluid with $\gamma=1$, we realize
that the emergence of finite-distance singularities persists even for this more
general type of bulk matter. A next step should therefore be to look for
possible ways to rectify these singularities.

In \cite{ack4}, we explored the possibility of avoiding the
singularities by constructing a regular matching solution from
(\ref{solution a-linear-flat}) and (\ref{solution rho-linear-flat}).
The procedure we followed there, is similar to the one used in \cite{forste3}:
we cut and match the part of the bulk that is free from finite-distance
singularities. This is indeed possible for an appropriate choice of the range
of parameters. In particular, we have examined the following two choices:
\begin{itemize}
\item [$I)$] $\gamma <-1$, $c_{2}\leq 0$,
with the $+$ sign for $Y<0$ and the $-$ sign for $Y>0$.

The matching solution then reads
\be
\label{solution a_g<-1}
a=\left(2(\gamma+1)\left(-\sqrt{\dfrac{2A c_{1}}{3}}|Y|+c_{2}\right)\right)^{1/(2(\gamma+1))},
\ee
and
\be
\label{solution rho_g<-1}
\rho=c_{1}\left(2(\gamma+1)\left(-\sqrt{\dfrac{2A c_{1}}{3}}|Y|+c_{2}\right)\right)^{-2},
\ee
with the brane placed at the origin $Y=0$. Clearly then, both $a$ and $\rho$
are non-singular since the term
$$\left(-\sqrt{2A c_{1}/3}|Y|+c_{2}\right)$$
is always negative.

\item [$II)$] $\gamma >-1$, $c_{2}\geq 0$, 
with the $+$ sign for $Y>0$ and the $-$ sign for $Y<0$.
Then the matching solution is
\be
a=\left(2(\gamma+1)\left(\sqrt{\dfrac{2A c_{1}}{3}}|Y|+c_{2}\right)\right)^{1/(2(\gamma+1))},
\ee
and
\be
\rho=c_{1}\left(2(\gamma+1)\left(\sqrt{\dfrac{2A c_{1}}{3}}|Y|+c_{2}\right)\right)^{-2}.
\ee
Again, $a$ and $\rho$ are non-singular, since the term
$$\left(\sqrt{2A c_{1}/3}|Y|+c_{2}\right)$$
is always positive.
\end{itemize}
We are going to focus only on the solution described by
(\ref{solution a_g<-1})-(\ref{solution rho_g<-1}), which corresponds to
$\gamma<-1$, since it is the only one of the two possibilities that leads to a finite
four-dimensional Planck mass.

To examine further the adequacy of this solution, we ought to check the
boundary conditions that describe the embedding of the brane in the bulk. A
natural condition to impose is, that the warp factor and energy density are continuous functions.
Note that in what follows, by writing $c_{i}^{+}$ ($c_{i}^{-}$) we refer to the value
of an arbitrary constant $c_{i}$ at $Y>0$ ($Y<0$). The continuity of the warp
factor at $Y=0$ leads to the condition
\be
(2(\gamma+1)c_{2}^{+})^{1/(2(\gamma+1))}=(2(\gamma+1)c_{2}^{-})^{1/(2(\gamma+1))},
\ee
or, since $c_{2}^{+}$ and $c_{2}^{-}$ are real numbers, we have
\be
\label{c2}
c_{2}^{+}=\pm c_{2}^{-},
\ee
depending on the value of $\gamma$.
Similarly, continuity of the density gives
\be
\dfrac{c_{1}^{+}}{(c_{2}^{+})^{2}}=\dfrac{c_{1}^{-}}{(c_{2}^{-})^{2}},
\ee
and using (\ref{c2}) we find
\be
\label{c1}
c_{1}^{+}=c_{1}^{-}.
\ee
On the other hand, the jump of the extrinsic curvature
$K_{\alpha\beta}=1/2(\partial g_{\alpha\beta}/\partial Y)$ ($\alpha,\beta=1,2,3,4$), is given by
\be
\label{junction}
K_{\alpha\beta}^{+}-K_{\alpha\beta}^{-}=-\kappa_{5}^{2}\left(S_{\alpha\beta}-
\dfrac{1}{3}g_{\alpha\beta}S \right),
\ee
where the surface energy-momentum tensor $S_{\alpha\beta}$ (defined only on the brane
and vanishing off the brane) is taken to be
\be
\label{surface tensor}
S_{\alpha\beta}=-g_{\alpha\beta}f(\rho),
\ee
with $f(\rho)$ denoting the brane tension and $S=g^{\alpha\beta}S_{\alpha\beta}$ the
trace of $S_{\alpha\beta}$. For our type of geometry, (\ref{junction}) becomes
\be
\label{junction a'}
a'(0^{+})-a'(0^{-})=-\dfrac{\kappa_{5}^{2}}{3}f(\rho(0))a(0).
\ee
Substitution of (\ref{solution a_g<-1}) and (\ref{surface tensor})
in (\ref{junction a'}), leads to a junction condition for the arbitrary constants
\be
\label{j1}
\sqrt{c_{1}}\left(\dfrac{1}{c_{2}^{+}}+\dfrac{1}{c_{2}^{-}}\right)=
 4\sqrt{\dfrac{2 A}{3}}(\gamma+1)f(\rho(0)),
 \ee
from which we see that we have to choose the plus sign in (\ref{c2}) and then (\ref{j1})
becomes
\be\label{ic}
\dfrac{\sqrt{c_{1}}}{c_{2}}=2\sqrt{\dfrac{2 A}{3}}(\gamma+1)f(\rho(0)).
\ee
\subsubsection{Energy conditions}
So far, we were able to construct a regular matching solution for
$\gamma<-1$, however, we still have to check if this range of $\gamma$ is
compatible with the energy conditions.

We begin with the weak energy condition: Keeping (\ref{wec_p}) as it is and
inputting $p=\gamma\rho$ in (\ref{wec}) gives
\be
p\geq 0 
\ee
and
\be
\label{wec-g}
(\gamma+1)\rho\geq 0
\ee
so that either
\be
\label{nec-linear 0}
\gamma\geq-1 \quad \textrm{and},\quad \rho\geq 0, \quad \textrm{or}, \quad \gamma\leq-1
\quad \textrm{and} \quad \rho\leq 0
\ee
For a flat brane, $\rho\geq 0$ because of (\ref{syst2i-linear-flat}), so we end up with
$\gamma>-1$ which is further refined to $\gamma>0$ upon the requirement of
(\ref{wec_p}). We end up with the condition
\be
\label{wec-linear-flat}
p\geq 0 \quad \textrm{and}\quad  \gamma>0.
\ee

For the strong energy condition, on the other hand, we input
$p=\gamma\rho$ in (\ref{sec_1}) and (\ref{sec_2}). We find that
\be
(-\gamma+1)\rho\geq 0 \quad \textrm{and}\quad (\gamma+1)\rho\geq 0
\ee
so
$$-1\leq\gamma\leq 1,$$ from which we find that it is possible to have either
$p=0$, or,
\be
\label{sec1-linear-flat}
p< 0 \quad \textrm{and}\quad -1\leq\gamma<0,
\ee
or,
\be
\label{sec2-linear-flat}
p> 0 \quad \textrm{and}\quad 0<\gamma\leq 1.
\ee

Finally the null energy condition (\ref{nec}) for $p=\gamma\rho$ gives
(\ref{wec-g}), which again implies that
\be
\label{nec-g-linear-flat}
\gamma\geq -1.
\ee
The inequalities (\ref{wec-linear-flat}), (\ref{sec1-linear-flat}), (\ref{sec2-linear-flat}) and
(\ref{nec-g-linear-flat}) show that the energy conditions restrict $\gamma$ to be at least
greater than or equal to $-1$, which means that the regular solution for $\gamma<-1$, cannot satisfy
the energy conditions.
\subsubsection{Planck Mass}
In this Section, we will show that the solution (\ref{solution a_g<-1}),
provides the appropriate range of $\gamma$ to obtain a finite
four-dimensional Planck mass.

The value of the four-dimensional Planck mass, $M_{p}^{2}=8\pi/\kappa$, is
determined by the following integral \cite{forste3}
\be
\label{planck mass}
\frac{\kappa_{5}^{2}}{\kappa}=\int_{-Y_{c}}^{Y_{c}}a^{2}(Y)dY.
\ee
For our solution, Eq.~(\ref{solution a_g<-1}), the above integral becomes \cite{ack4},
\beq
& &\int_{-Y_{c}}^{Y_{c}}
\left(2(\gamma+1)\left(-\sqrt{\dfrac{2 A c_{1}}{3}}|Y|+c_{2}\right)\right)^{1/(\gamma+1)}dY
=\\
&=&\dfrac{1}{2(\gamma+2)}\sqrt{\dfrac{3}{2A c_{1}}}\left(2(\gamma+1)\left(\sqrt{\dfrac{2Ac_{1}}{3}}Y
+c_{2}\right)\right)^{(\gamma+2)/(\gamma+1)}|_{-Y_{c}}^{0}-\\
&-&\dfrac{1}{2(\gamma+2)}\sqrt{\dfrac{3}{2A c_{1}}}\left(2(\gamma+1)\left(-\sqrt{\dfrac{2Ac_{1}}{3}}Y
+c_{2}\right)\right)^{(\gamma+2)/(\gamma+1)}|_{0}^{Y_{c}},
\eeq
In the limit $Y_{c}\rightarrow\infty$, we see that the Planck mass remains finite only for
\be
\label{finite_planck}
-2<\gamma<-1,
\ee
and takes the form
\be
\frac{\kappa_{5}^{2}}{\kappa}=\sqrt{\dfrac{3}{2A c_{1}}}
\dfrac{(2(\gamma+1)c_{2})^{\frac{\gamma+2}{\gamma+1}}}{
\gamma+2}.
\ee
This means that the interval $(-\infty,-1)$ for which the solution (\ref{solution a_g<-1})
is defined, has to be refined to $(-2,-1)$, after taking into account the requirement of a finite Planck mass.
Combining this fact with the results of the previous subsection, we
conclude that for a flat brane and a linear fluid with $\gamma\neq -1$, it is not feasible to construct a
regular solution that satisfies both the requirement for a finite Planck mass
given by (\ref{finite_planck}) \emph{and} the energy conditions.
\subsection{The special case $\gamma=-1$ for a flat brane}
Putting $\gamma=-1$ in (\ref{syst2iii-linear}), we find
\be
\label{rho gamma=-1}
\rho=c_{1},
\ee
where $c_{1}$ is an integration constant. Since $\rho\geq 0$ from (\ref{syst2i-linear-flat}),
we see that $c_{1}$ has to be non-negative. Substituting (\ref{rho gamma=-1}) in
(\ref{syst2i-linear-flat}) we find
\be
\label{a gamma=-1}
a(Y)=e^{\pm\sqrt{({\kappa_{5}}^{2}/6) c_{1}}Y+c_{2}},
\ee
where $c_{2}$ is an integration constant. We note that this solution has no finite distance
singularities and satisfies trivially the null energy condition ($\gamma=-1$). For a finite four-dimensional
Planck mass, we can make the following choice: we can choose the $+$ sign for $Y<0$
and the $-$ sign for $Y>0$ and place the brane at $Y=0$. Then the matching solution reads
\be
\label{a match gamma=-1}
a(Y)=e^{-\sqrt{({\kappa_{5}}^{2}/6)c_{1}}|Y|+c_{2}}.
\ee
This reduces to the Randall-Sundrum solution of \cite{rs2}, by setting
$c_{2}=0$ and $c_{1}=-\Lambda$, where $\Lambda< 0$ is the bulk
cosmological constant in that model.

Continuity of the warp factor and density at the position of the brane give
\be
c_{2}^{+}=c_{2}^{-}, \quad \textrm{and} \quad c_{1}^{+}=c_{1}^{-}.
\ee
For simplicity, we can also set $c_{2}=0$. Then using the junction condition
(\ref{junction a'}), we can find the form of the brane tension which reads
\be
f(\rho(0))=\frac{\sqrt{6c_{1}}}{\kappa_5},
\ee
and we note that the tension is positive.

Finally, the four-dimensional Planck mass is determined by (\ref{planck mass}). Here we have
\be
a^{2}(Y)=e^{-2\sqrt{({\kappa_{5}}^{2}/6)c_{1}}|Y|},
\ee
and using the symmetry of the solution we find
\be
\frac{\kappa_{5}^{2}}{\kappa}=\int_{-Y_{c}}^{Y_{c}}a^{2}(Y)dY=2\int_{0}^{Y_{c}}
e^{-2\sqrt{({\kappa_{5}}^{2}/6)c_{1}}Y}dY=-\frac{\sqrt{6}}{\kappa_{5}\sqrt{c_{1}}}
e^{-2\sqrt{({\kappa_{5}}^{2}/6)c_{1}}Y}|_{0}^{Y_{c}}.
\ee
Taking $Y_{c}\rightarrow \infty$, we see that the four-dimensional Planck mass remains
finite and is proportional to
\be
\frac{\sqrt{6}}{\kappa_{5}\sqrt{c_{1}}}.
\ee
\subsection{Curved brane}
The impossibility of finding for a range of $\gamma$ a flat-brane solution bearing the required
physical properties mentioned in the previous subsections, led us to
further research the question of whether the situation could be resolved
by allowing for a nonzero brane curvature. Of course, this would inevitably
bring back the cc-problem, as mentioned in the introduction. Still, it is worth
exploring the impact of the curvature, as this would offer a deeper
understanding of the factors that monitor the dynamics and evolution of the
brane-worlds under investigation.

Assuming $k\neq 0$ in Eqs.~(\ref{syst2i-linear}) and (\ref{syst2ii-linear})
and substituting (\ref{rho to a-linear}) in (\ref{syst2i-linear}), we find
\be
\label{syst2i-linear-curved}
a'^2=\dfrac{2}{3}A  c_{1}a^{-2(2\gamma+1)}+kH^2.
\ee
For simplicity, we can set $C=2/3Ac_{1}$ and keep in mind that the sign of $C$
follows the sign of $\rho$. Then (\ref{syst2i-linear-curved}) can be written as
\be
\label{integration eq}
a'^2=Ca^{-2(2\gamma+1)}+kH^2.
\ee

It automatically follows from (\ref{integration eq}) that the LHS of
this equation restricts the acceptable combinations of the signs of $C$ and
$k$ and the possible asymptotic behaviors of $a$. For example, the case
$C<0$ and $k<0$ becomes an impossible combination. Also, the case
$C<0$, $k>0$ becomes possible only for,
\begin{equation}
\label{bound ds}
\textrm{dS brane-world:}\quad
0<a^{-2(2\gamma+1)}<-\dfrac{kH^2}{C},
\end{equation}
while, the case $C>0$, $k<0$ is allowed only for
\begin{equation}
\label{bound ads}
\textrm{AdS brane-world:}\quad a^{-2(2\gamma+1)}>-\dfrac{kH^2}{C}>0.
\end{equation}

Clearly, the above inequalities show that both cases are likely to give rise
to regular solutions. In particular, for the case $C<0$ and a dS brane with
$\gamma>-1/2$, (\ref{bound ds}) implies that
$a^{2(2\gamma+1)}>-C/(kH^{2})>0$, so that the warp factor, $a$, is bounded
away from zero, which prevents collapse singularities from happening.
The only way that this case may introduce a finite-distance singularity, is by
allowing the warp factor to become divergent within a finite distance, thus signalling a big-rip singularity.
However, further calculations presented in \cite{ack6}, showed that this
singular behaviour is also excluded, which means that for $C<0$ and a dS brane there
is indeed a regular solution.

On the other hand, for the case $C>0$ and an AdS brane with
$\gamma<-1/2$, (\ref{bound ads}) implies that the warp factor
is again bounded away from zero, thus excluding the existence of collapse
singularities here, as well.  However, for this latter case, we have to further
restrict $\gamma$ on the interval $(-1,-1/2)$, in order to avoid a finite-distance big-rip
singularity \cite{ack6}.

To derive solutions for a curved brane, we can proceed by writing
Eq. (\ref{integration eq}) in the form,
\be
\label{hyper-linear}
\int \frac{da}{\sqrt{Ca^{-2(2\gamma+1)}+kH^{2}}}=\pm \int dY.
\ee
Naturally, the integration is more complicated in this case and cannot be
performed directly. We can nevertheless, express our solutions implicitly
by the Gaussian hypergeometric function $_{2}F_{1}(\alpha,b,c;z)$, defined by
\be
_{2}F_{1}(\alpha,b,c;z)=\frac{\Gamma(c)}{\Gamma(\alpha)\Gamma(b)}
\sum_{n=0}^{\infty}\frac{\Gamma(\alpha+n)\Gamma(b+n)}{\Gamma(c+n)}\frac{z^{n}}{n!}
\ee
and convergent for $|z|<1$, where $\Gamma$ is the Gamma function.
For this purpose, we use standard substitution formulas and bring the integral
on the LHS of Eq.~(\ref{hyper-linear}) in the form of an integral representation
of a hypergeometric function $_{2}F_{1}(\alpha,b,c;z)$ given by \cite{wang}
\be
\label{integral theory}
_{2}F_{1}(\alpha,b,c;z)=\frac{\Gamma(c)}{\Gamma(b)\Gamma(c-b)}\int_{0}^{1}
t^{b-1}(1-t)^{c-b-1}(1-tz)^{-\alpha}dt
\ee
and provided that
\be
\label{hyper cond1}
0<Re(b)<Re(c).
\ee
In our solutions, the corresponding parameters $b$ and $c$ are, either, constants
or, functions of $\gamma$. This means that condition (\ref{hyper cond1}) above,
determines the range of $\gamma$, for which the representations of solutions
in terms of hypergeometric functions are valid.

We give a full list of solutions derived, in this way, below.

\begin{itemize}
\item [$I)$] For a dS brane, $C>0$ and
\item[$Ia)$] $\gamma<-1/2$, the solution is,
\be
\label{sol type Ia}
\pm Y+C_{2}=\dfrac{a}{\sqrt{kH^2}}\,_{2}F_{1}\left(\dfrac{1}{2},-\dfrac{1}{2(2\gamma+1)},
\dfrac{4\gamma+1}{2(2\gamma+1)};-\dfrac{C}{kH^2}a^{-2(2\gamma+1)}\right),
\ee
where $C_{2}$ is an integration constant.

\item[$Ib)$] $\gamma>-1/2$, the solution reads,
\be
\label{sol type Ib}
\pm Y+C_{2}=\dfrac{1}{2(\gamma+1)\sqrt{C}}a^{2(\gamma+1)}\,
_{2}F_{1}\left(\dfrac{1}{2}, \dfrac{\gamma+1}{2\gamma+1},\dfrac{3\gamma+2}{2\gamma+1};
-\dfrac{k H^2}{C}a^{2(2\gamma+1)}\right).
\ee

\item [$II)$] For dS brane, $C<0$ and
\item [$IIa)$] $\gamma<-1/2$, the solution is given
by (\ref{sol type Ia}).

\item [$IIb)$] $\gamma>-1/2$, the solution is,
\beq
\nonumber
\label{sol type IIb}
\pm Y +C_{2}&=&\dfrac{(-C)^{-\frac{\gamma}{2\gamma+1}}(kH^2)^{\frac{-1}{2(2\gamma+1)}}}
{2(2\gamma+1)}\sqrt{a^{2(2\gamma+1)}+\dfrac{C}{kH^2}}\times\\
& &\times _{2}F_{1}\left(\dfrac{1}{2},\dfrac{\gamma}{2\gamma+1},\dfrac{3}{2};
1+\dfrac{k H^2}{C}a^{2(2\gamma+1)}\right).
\eeq
\item[$III)$] For AdS, $C>0$ and
\item[$IIIa)$] $\gamma>-1/2$, or, $\gamma<-1$ the solution is
given by Eq. (\ref{sol type Ib}).

\item[$IIIb)$] $-1<\gamma<-1/2$ the solution is,
\beq
\nonumber
\label{sol type IIIb}
\pm Y+C_{2}&=&-\dfrac{C^{\frac{\gamma+1}{2\gamma+1}}}{2(2\gamma+1)(-k H^2)^
{\frac{4\gamma+3}{4\gamma+2}}}
\sqrt{a^{-2(2\gamma+1)}+\dfrac{kH^2}{C}}\,\times\\
&\times&_{2}F_{1}
\left(\dfrac{4\gamma+3}{2(2\gamma+1)},\dfrac{1}{2}, \dfrac{3}{2};
1+\dfrac{C}{k H^2}a^{-2(2\gamma+1)}\right).
\eeq
\end{itemize}

We can deduce the asymptotic behaviors of the warp factor either directly, or,
indirectly from the above implicit solutions. For example, take the case $Ia)$
that is valid for a dS brane with $C>0$ and $\gamma<-1/2$: the argument of the
hypergeometric function is
$$z=-\dfrac{C}{kH^2}a^{-2(2\gamma+1)}.$$
Note that, the power of $a$ is positive and so,
letting $a\rightarrow 0$, makes $z\rightarrow 0$ which
means that the hypergeometric function is convergent and that it, actually,
approaches one. Then from the LHS of (\ref{sol type Ia}), we see that $Y$ will
approach the finite value $\pm C_{2}$, which shows that this solution has
a finite-distance collapse singularity.

A more indirect case is the one described by $IIIb)$. As it follows from
(\ref{bound ads}), the warp factor is bounded away from zero so that
no collapse singularities exist in this case. We should, however, check if it is
possible to have a divergent warp factor within finite distance. Since the
power of $a$ in the argument
$$z=1+\dfrac{C}{k H^2}a^{-2(2\gamma+1)}$$
of the hypergeometric function is positive, letting
$a\rightarrow\infty$ leads us outside the disc of convergence of $_{2}F_{1}$.
In order to proceed and find the behavior of the solution
as $a\rightarrow\infty$, we use the following rule \cite{wang}
\beq
\nonumber
_{2}F_{1}(\alpha,b,c;z)&=&A (-z)^{-\alpha}\,_{2}F_{1}(\alpha,\alpha-c+1,
\alpha-b+1;1/z)+\\
\label{hyper-infty}
&+&B (-z)^{-b}\,_{2}F_{1}(b,b-c+1,b-\alpha+1;1/z),
\eeq
where the constants $A$ and $B$ are given by
\be
A=\frac{\Gamma(c)\Gamma(b-\alpha)}{\Gamma(c-\alpha)\Gamma(b)}
\ee
and
\be
B=\frac{\Gamma(c)\Gamma(\alpha-b)}{\Gamma(c-b)\Gamma(\alpha)}.
\ee

Using (\ref{hyper-infty}), the hypergeometric function of solution
$IIIb)$ transforms to
\beq
\nonumber
& &_{2}F_{1}\left( \dfrac{4\gamma+3}{2(2\gamma+1)},\dfrac{1}{2},\dfrac{3}{2},
1+\dfrac{C}{k H^2}a^{-2(2\gamma+1)}\right)=\\
\nonumber
& &A_{1}\left(-1-\dfrac{C}{kH^2}a^{-2(2\gamma+1)}\right)^{-\frac{(4\gamma+3)}{4\gamma+2}}
\times\\
\nonumber
\label{hyper aux}
& &\times_{2}F_{1}\left(\dfrac{4\gamma+3}{2(2\gamma+1)}, \dfrac{\gamma+1}{2\gamma+1},
\dfrac{3\gamma+2}{2\gamma+1}; \left(1+\dfrac{C}{kH^2}a^{-2(2\gamma+1)}\right)^{-1}\right)+\\
\nonumber
& & +B_{1}\left(-1-\dfrac{C}{kH^2}a^{-2(2\gamma+1)}\right)^{-1/2}\times\\
& &\times_{2}F_{1}\left(\dfrac{1}{2},0,\dfrac{\gamma}{2\gamma+1};
\left(1+\dfrac{C}{kH^2}a^{-2(2\gamma+1)}\right)^{-1}\right),
\eeq
where $A_{1}$ and $B_{1}$ are constants given by,
\be
A_{1}=\dfrac{\Gamma(-(\gamma+1)/(2\gamma+1))}{2\Gamma(\gamma/(2\gamma+1))},
\ee
and
\be
B_{1}=\dfrac{\sqrt{\pi}\,\Gamma((\gamma+1)/(2\gamma+1))}
{2\Gamma((4\gamma+3)/(2(2\gamma+1)))}.
\ee
Substituting (\ref{hyper aux}) in (\ref{sol type IIIb}), we
deduce that $a$ behaves according to (we denote this below,
with the symbol $\sim$) \cite{ack6}
\be\label{asyminfty}
a^{2(\gamma+1)}\sim\pm Y+C_{2},
\ee
and so, letting $a\rightarrow\infty$, gives $Y\rightarrow\pm\infty$.
This means that the solution (\ref{sol type IIIb}) is indeed regular.

We can use the same procedure for determining the singular, or, regular nature of
solutions.
For a convenient overall view of the most important behaviors of
the solutions of cases $Ia)$-$IIIb)$, we use the table below to illustrate them.
Each behavior is accompanied with information regarding the corresponding
range of $\gamma$ and the solution from which it arises.
For brevity, regular behaviors like $a\rightarrow\infty$ as
$Y\rightarrow\infty$, that coexist with finite-distance singularities
are not depicted in the table but can be found in \cite{ack6}.

\begin{table}[!h]
\caption{Asymptotic behaviors for a curved brane and a linear fluid}
\label{table_example}
\begin{tabular}{l|l|l|l|l}
\hline 
$\gamma$ & $(-\infty,-1)$ & $(-1,-1/2)$ & $(-1/2,1/2) $ & $(1/2,\infty)$\\ \hline \hline
dS, $C>0$ & collapse sing., $Ia$ & collapse sing., $Ia$ & collapse sing., $Ib$ &
collapse sing., $Ib$\\
& big-rip sing., $Ia$ & & &\\ \hline
dS, $C<0$ & collapse sing., $IIa$  & collapse sing., $IIa$  & {\em
regular}, $IIb$ & {\em regular}, $IIb$ \\ \hline
AdS, $C>0$ & big-rip sing., $IIIa$ & {\em regular}, $IIIb$ & collapse sing.,
$IIIa$ & collapse sing., $IIIa$ \\\hline
\end{tabular}
\vspace*{-4pt}
\end{table}


Summarizing, for a curved brane we obtain regular solutions
from the following two cases:
\begin{itemize}
\item $IIb)$, referring to a dS brane with negative density and $\gamma> -1/2$

\item $IIIb)$, referring to an AdS brane with positive density and $-1<\gamma<-1/2$.
\end{itemize}

We note here that the special case of $\gamma=-1$, is studied separately
in a subsequent subsection.
\subsubsection{The null energy condition}
The case of curved branes, has already proved successful in providing regular
solutions, which was an impossible outcome for a flat brane. We still have to check
whether the regular solutions can fulfil energy conditions as well as the requirement of a
finite Planck mass.

For a linear fluid, we have already seen that the null energy condition is given by (\ref{nec-linear 0}). For a curved brane it
translates to having
\be
\rho\geq 0 \quad \textrm{and} \quad \gamma\geq-1,\quad\textrm{or,}\quad
\rho\leq 0 \quad \textrm{and}\quad \gamma\leq-1.
\ee
These two conditions may be written equivalently with
respect to $C$ instead of $\rho$ as,
\be
\label{nec_c_1}
C\geq 0 \quad \textrm{and}\quad  \gamma\geq-1,
\ee
and
\be
\label{nec_c_2}
C\leq 0 \quad \textrm{and} \quad \gamma\leq-1.
\ee

Combining conditions (\ref{nec_c_1}) and (\ref{nec_c_2}) with the ranges of
$\gamma$ and $C$ for which the regular solutions of cases $IIb)$ and $IIIb)$
are defined, we see that only the regular solution of $IIIb)$ is compatible
with the null energy condition.
\subsubsection{Localisation of gravity}
To complete our study for a curved brane, we examine in this subsection, the
requirement of a finite Planck mass.

We continue, for illustration purposes, to focus on the regular solution of
case $IIIb)$. The 4D-Planck mass is given by
the integral of Eq.~(\ref{planck mass}). The behavior of $a^{2}$ that
we need to substitute in Eq.~(\ref{planck mass}), can be deduced from
(\ref{asyminfty})
\be
\label{asq gamma}
a^{2}\sim(|Y|+c_{2})^{\frac{1}{\gamma+1}},
\ee
after setting $\pm Y=|Y|$ and positioning the brane
at $Y=0$ \cite{ack6}. It is straightforward to see that integration of $a^{2}$ gives an
expression with $Y$ raised to the exponent,
\be
\dfrac{\gamma+2}{\gamma+1},
\ee
which is positive, since $-1<\gamma<-1/2$ for the case $IIIb)$.
Therefore, the Planck mass is infinite in this case. We note that for a
finite Planck mass, we need to have
$$-2<\gamma<-1.$$

As shown in \cite{ack6}, also the second regular solution of case $IIb)$ 
fails to give a finite Planck mass, as well. The problem of localizing gravity
on the brane, persists further in the case of regular matching solutions that can be constructed out of the singular solutions $Ia)$ and $Ib)$.

Summarizing, for the case of a curved brane and a linear fluid with
$\gamma\neq -1$, there exist regular solutions for ranges of $\gamma$, however these ranges are
inconsistent, with the requirement of a finite Planck mass (AdS brane with positive density),
or, with both the null energy condition and the requirement of a finite Planck
mass (dS brane with negative density). On the other hand, regular matching
solutions that can be constructed by cutting the bulk and gluing the parts that are free from singularities,
satisfy the null energy conditions but fail to localize gravity on the brane \cite{ack6}.
\subsection{The special case $\gamma=-1$ for a curved brane}
For $\gamma=-1$ and a curved brane, we find from (\ref{syst2iii-linear}), that
\be
\label{rho gamma=-1 curved}
\rho=c_{3},
\ee
where $c_{3}$ is an integration constant.
Substituting (\ref{rho gamma=-1 curved}) and
$\gamma=-1$ in (\ref{syst2ii-linear}) we find
\be
a''-\kappa_{5}^{2}\dfrac{c_{3}}{6}a=0.
\ee
For $c_{3}>0$ the above equation has
the general solution 
\be
\label{a gamma=-1 curved}
a(Y)=c_{1}e^{{\kappa_{5}}\sqrt{c_{3}/6}Y}+c_{2}e^{{-\kappa_{5}}\sqrt{c_{3}/6}Y}, 
\ee
where $c_{1}$ and $c_{2}$ are arbitrary constants. Substitution of (\ref{a gamma=-1 curved})
in (\ref{syst2i-linear}), determines the arbitrary constant $c_{3}$ in terms of $c_{1}$ and $c_{2}$.
Here we find
\be
c_{3}=-\dfrac{3 kH^{2}}{2c_{1}c_{2}\kappa_{5}^{2}}.
\ee
Since $c_{3}>0$ we need to have the following restrictions
on the signs of $c_{1}$, $c_{2}$ and $k$
\be
\textrm{either} \quad c_{1}c_{2}<0 \quad \textrm{and} \quad k>0, \quad \textrm{or,}\quad c_{1}c_{2}>0 \quad \textrm{and} \quad k<0.
\ee
For $c_{1}c_{2}<0$ and $k>0$, there is a finite-distance singularity at
\be
Y_{s}=\dfrac{\sqrt{6}}{2\kappa_{5}\sqrt{c_{3}}}\ln \left(-\dfrac{c_{2}}{c_{1}}\right).
\ee
For $c_{1}c_{2}>0$ and $k<0$, on the other hand, the above singularities are excluded.
Also, from (\ref{a gamma=-1 curved}), it follows that both $c_{1}$ and $c_{2}$
have to be positive. This is exactly why the warp factor cannot approach zero
within finite distance. Since divergence of the warp factor within finite
distance is also impossible, we conclude that this solution is regular.
In addition, we note that this solution satisfies the null energy condition trivially ($\gamma=-1$).
Next we introduce a brane, examine the boundary conditions at the
position of the brane and comment on the resulting four-dimensional Planck mass.

As before, we can place the brane at $Y=0$ and construct the matching solution
\be
\label{k<0 gamma=-1}
a(Y)=c_{1}e^{\sqrt{H^{2}/(4c_{1}c_{2})}|Y|}+c_{2}e^{-\sqrt{H^{2}/(4c_{1}c_{2})}|Y|}.
\ee
Then we can express the conditions imposed by the continuity of the warp
factor and density at the position of the brane, in terms of the arbitrary
constants $c_{1}$ and $c_{2}$, we find
\be
c_{1}^{+}+c_{2}^{+}=c_{1}^{-}+c_{2}^{-}\quad \textrm{and}\quad
c_{1}^{+}c_{2}^{+}=c_{1}^{-}c_{2}^{-},
\ee
which implies that $c_{1}^{+}=c_{1}^{-}=c_{1}$ and $c_{2}^{+}=c_{2}^{-}=c_{2}$.
Using the junction condition (\ref{junction a'}), we can find the form of the
brane tension, it reads
\be
f(\rho(0))=\frac{3}{{\kappa_5}^{2}}\sqrt{\frac{H^{2}}{c_{1}c_{2}}}\frac{c_{2}-c_{1}}{c_{1}+c_{2}}.
\ee
We note that the sign of the tension depends on the ordering between
$c_{1}$ and $c_{2}$: for $c_{2}>c_{1}$ the tension is positive, while for
$c_{2}<c_{1}$ the tension is negative.

Finally, we see that for the solution (\ref{k<0 gamma=-1}), the integral
in (\ref{planck mass}) that determines the four-dimensional Planck mass,
diverges. The only way of ending up with a finite Planck mass from
this solution is by compactifying $Y$ with a second brane.
\section{Non-linear fluid}
The problems we faced with the solutions of flat, or, curved branes analysed
above, can be resolved by allowing $\lambda\neq 1$, in the equation of state
(\ref{eos}).
We are going to briefly review here, the main features of solutions
for a non-linear equation of state and a flat brane, while full details can be found in
\cite{ack7}.

We start by substituting $k=0$ in the system (\ref{syst2i})-(\ref{syst2iii}), 
leading to
\bq
\label{syst2i-nonlinear}
\frac{a'^{2}}{a^{2}}&=&\frac{\kappa_{5}^{2}}{6}\rho,\\
\label{syst2ii-nonlinear}
\frac{a''}{a}&=&-\frac{\kappa_{5}^{2}}{6}{(2\gamma\rho^{\lambda}+\rho)},
\eq
and
\be
\label{syst2iii-nonlinear}
\rho'+4(\gamma\rho^{\lambda}+\rho)\frac{a'}{a}=0.
\ee

Before solving the system of equations above, we can first check the
restrictions that the null energy condition imposes on the parameters of
a non-linear fluid. Inputting $p=\gamma\rho^{\lambda}$ in the null energy
condition (\ref{nec}), we obtain
\be
\label{nec-0}
\gamma\rho^{\lambda}+\rho\geq 0,
\ee
or, equivalently,
\be
\rho^{\lambda}(\gamma+\rho^{1-\lambda})\geq 0.
\ee
Since $\rho\geq 0$ from Eq.~(\ref{syst2i-nonlinear}), we see that the
null energy condition can be written as
\be
\label{nec-nonlinear}
\gamma+\rho^{1-\lambda}\geq 0.
\ee

To derive a solution of the system of
Eqs.~(\ref{syst2i-nonlinear})-(\ref{syst2iii-nonlinear}), we integrate the
continuity equation (\ref{syst2iii-nonlinear}) to find the
relation between the warp factor and the density. In the integration process
we arrive at a logarithmic term of the form
$\ln |\gamma+\rho^{1-\lambda}|$. To incorporate from the beginning the null
energy condition (\ref{nec-nonlinear}), we choose to ignore the absolute value and simply
put this term equal to $\ln (\gamma+\rho^{1-\lambda})$. The resulting relation
between $\rho$ and $a$ is
\be
\label{rho to a-nonlinear}
\rho={(-\gamma+c_{1}a^{4(\lambda-1)})}^{1/(1-\lambda)},
\ee
where
\be c_1=\frac{\gamma+\rho_0^{1-\lambda}}{a_0^{4(\lambda-1)}},
\ee
with $\rho_0=\rho(Y_0), a_0=a(Y_0)$ being the initial conditions.
According to (\ref{nec-nonlinear}) this translates to $c_{1}\geq 0$.

To avoid the singularity in the density with $\rho\rightarrow\infty$
for $\lambda>1$ and
\be
\label{sing}
a^{4(\lambda-1)}=\dfrac{\gamma}{c_{1}},
\ee
we take, in what follows, $\gamma<0$.

Next, we substitute (\ref{rho to a-nonlinear}) in (\ref{syst2i-nonlinear})
and integrate. We find
\be
\label{hyper_integral-nonlinear}
\int\dfrac{a}{(c_{1}-\gamma a^{4(1-\lambda)})^{1/(2(1-\lambda))}}\,da=\pm\dfrac{\kappa_{5}}{\sqrt{6}}\int dY.
\ee
We note that we can calculate directly the above integral for values
of $\lambda$ that make $1/(2(1-\lambda))$ a negative integer. These are:
$\lambda=(n+1)/n$, with $n=2k$ and $k$ a positive integer. We study a
characteristic example of such choice in the next paragraph.
\subsection{The case of $\lambda=3/2$}
Let us focus first on the simplest case, $n=2$ corresponding to $\lambda=3/2$.
This makes the exponent $1/(2(1-\lambda))$ in the integral on the LHS of
Eq.~(\ref{hyper_integral-nonlinear}), equal to $-1$. It is then
straightforward to integrate. We arrive at the following implicit solution
\be
\label{sol_3/2}
\pm Y+C_{2}=\dfrac{\sqrt{6}}{\kappa_{5}}\left(\dfrac{c_{1}}{2} a^{2}-\gamma\ln a\right),
\ee
where $C_{2}$ is an integration constant. Looking at solution (\ref{sol_3/2}), we
see that we can have the following asymptotic behaviors
\beq
\label{a_div_gen}
a&\rightarrow&\infty, \quad\rho\rightarrow  0,\quad p\rightarrow  0, \quad \textrm{as}\quad Y\rightarrow\pm\infty\\
\label{a_0_gen}
a&\rightarrow&0^{+}, \quad  \rho\rightarrow 1/\gamma^{2},\quad p\rightarrow -1/\gamma^{2},\quad \textrm{as}\quad Y\rightarrow\pm\infty,
\eeq
which show that all pathological behaviors of $a$, 
become possible only at infinite distance, and therefore this solution is regular.

In addition to its good features of regularity and compatibility with the null
energy condition, the solution (\ref{sol_3/2}) also offers the possibility to
construct a matching solution that leads to a finite 4D-Planck mass; hence, it embodies all the required physical properties.
The matching solution reads \cite{ack7}
\be
\label{matching sol 3/2}
|Y|=\dfrac{\sqrt{6}}{\kappa_{5}}\left(-\dfrac{c_{1}}{2}a^{2}+\gamma\ln a-\dfrac{\gamma}{2}-\gamma \ln \sqrt{\dfrac{-\gamma}{c_{1}}}\right),\quad 0<a\leq\sqrt{\dfrac{-\gamma}{c_{1}}},
\ee
with the brane positioned at $Y=0$.

To calculate the 4D-Planck mass, 
we first figure out from (\ref{matching sol 3/2}) the behaviour of $a^{2}$ as $Y\rightarrow-\infty$, which reads
\be
\label{sq a sol 3/2}
a^{2}\sim e^{-(\sqrt{6}\kappa_{5}/(3\gamma))Y}.
\ee
Then by using the symmetry of (\ref{matching sol 3/2}), we write the integral
(\ref{planck mass}) in the following form
\be
\int_{-Y_{c}}^{Y_{c}}a^{2}(Y)dY=2\int_{-Y_{c}}^{0}a^{2}(Y)dY\sim
2\int_{-Y_{c}}^{0}e^{-(\sqrt{6}\kappa_{5}/(3\gamma))Y}dY=
-\sqrt{6}\dfrac{\gamma}{\kappa_{5}} (1-e^{(\sqrt{6}\kappa_{5}/(3\gamma))Y_{c}}).
\ee
Taking $Y_{c}\rightarrow\infty$ and keeping in mind that we consider only negative values
of $\gamma$, we see that the Planck mass remains finite and is proportional to
$$-\sqrt{6}\dfrac{\gamma}{\kappa_{5}}.$$
\subsection{Solutions for general $\lambda$}
For general $\lambda$, solving the system
(\ref{syst2i-nonlinear})-(\ref{syst2iii-nonlinear}) becomes much more
complicated. For a convenient overview, we outline below, the types of new solutions that we obtain
and comment on their asymptotic behaviors. Full details can be found in \cite{ack7}.

For all values of $\lambda>1$, the solutions share all the fine qualities,
previously, encountered in the solution for $\lambda=3/2$. In particular,
for $\lambda=1+1/(2k)$, with $k$ a positive integer, we find the following
form of solution
\be
\label{sol_endpoints}
\pm Y+c_{2}=\dfrac{\sqrt{6}}{\kappa_{5}}\left(\sum_{s=0}^{k-1}\dfrac{k!}{(k-s)!s!}\dfrac{c_{1}^{k-s}}{2-2s/k} a^{2-2s/k}(-\gamma)^{s}+(-\gamma)^{k}\ln a\right).
\ee
Furthermore, for $\lambda>3/2$ we have the solution
\beq
\label{sol_3/2 above}
\nonumber
\pm Y+c_{2}&=&\dfrac{\sqrt{6}}{\kappa_{5}}\left(\dfrac{a^{2}}{2}(c_{1}-\gamma a^{4(1-\lambda)})^{1/(2(\lambda-1))}-
 \dfrac{\gamma c_{1}^{(3-2\lambda)/(2(\lambda-1))}}
{2(3-2\lambda)}a^{2(3-2\lambda)}\times \right. \\
& &\left.
_{2}F_{1}\left(\dfrac{3-2\lambda}{2(1-\lambda)},\dfrac{3-2\lambda}{2(1-\lambda)},\dfrac{3-2\lambda}{2(1-\lambda)}+1;\dfrac{\gamma}{c_{1}}a^{4(1-\lambda)}\right)\right),
\eeq
For $1<\lambda<3/2$, on the other hand, we have solutions valid inside intervals
of the form $(1+1/(2k),1+1/2(k-1))$ with $k$ a positive integer such that $k\geq2$,
given by
\beq
\label{sol_n+1/n_n-1/n-2}
\nonumber
\pm Y+c_{2}&=&\frac{\sqrt{6}}{\kappa_{5}}\left(\sum_{s=0}^{n/2-1}\dfrac{(-\gamma)^{s}}{2(1-2s(\lambda-1))}\left(c_{1}a^{4(\lambda-1)}-\gamma\right)^{1/(2(\lambda-1))-s}\right.\\ \nonumber
&+&\left.\dfrac{(-\gamma)^{n/2}(c_{1})^{((n+1)-n\lambda)/(2(\lambda-1))}}{2((n+1)-n\lambda)}a^{2((n+1)-n\lambda)}\times \right.\\
&\times&\left._{2}F_{1}\left(\dfrac{(n+1)-n\lambda}{2(1-\lambda)},\dfrac{(n+1)-n\lambda}{2(1-\lambda)},\dfrac{(n+1)-n\lambda}{2(1-\lambda)}+1;\dfrac{\gamma}{c_{1}}a^{4(1-\lambda)}\right)\right).
\eeq

All solutions for $\lambda>1$ are free from finite-distance singularities, and
follow the asymptotic behaviors
\beq
\label{a_div}
a&\rightarrow&\infty, \quad \rho\rightarrow  0,\quad p\rightarrow  0, \quad \textrm{as}\quad Y\rightarrow\pm\infty\\
\label{a_0}
a&\rightarrow&0^{+}, \quad \rho\rightarrow(-\gamma)^{1/(1-\lambda)},\quad p\rightarrow -(-\gamma)^{1/(1-\lambda)},\quad \textrm{as}\quad Y\rightarrow\pm\infty.
\eeq
We can also construct a matching solution with a finite Planck
mass from every solution with $\lambda>1$, by applying the method
presented in the previous Sections \cite{ack7}.

Finally, for $\lambda<1$, the situation changes drastically, because
of the emergence of finite-distance singularities of the collapse type.
In particular, the solution for $\lambda<1$ reads
\be
\label{a_sing}
\pm Y+c_{2}=\frac{\sqrt{6}}{2\kappa_{5}}c_{1}^{1/(2(\lambda-1))}a^{2}\,_{2}F_{1}\left(\dfrac{1}{2(1-\lambda)},\dfrac{1}{2(1-\lambda)},
\dfrac{1}{2(1-\lambda)}+1;\dfrac{\gamma}{c_{1}}a^{4(1-\lambda)}\right).
\ee
As shown in \cite{ack7}, this solution has a finite-distance singularity at
$Y\rightarrow \pm c_{2}$, with
\beq
& & a\rightarrow 0^{+}, \quad \rho\rightarrow\infty, \quad p\rightarrow 0, \quad \textrm{if}\quad
\lambda<0\\
& & a\rightarrow 0^{+}, \quad \rho\rightarrow\infty, \quad p\rightarrow \infty,\quad \textrm{if}\quad
0<\lambda<1.
\eeq
The behaviors of $p$ and $\rho$ above, have been deduced from Eqs.~(\ref{eos})
and (\ref{rho to a-nonlinear}).
\section{Conclusions and open questions}
We have reviewed the effect of the curvature and equation of state of the bulk fluid
in the behavior of solutions of brane-worlds, consisting of a 3-brane embedded
in a five-dimensional bulk.

For a linear equation of state with $\gamma\neq -1$ and a flat brane, there is
always a finite-distance singularity. The types of singularity that arise in this
case are, the collapse type which is determined by a vanishing warp factor and a divergent
density and pressure (for $\gamma\neq 0$), or, a big-rip type that is signatured by a divergent
warp factor and also density and pressure. The avoidance
of such singularities becomes possible only after cutting and matching the part of the bulk that
is free from singularities. Still, the solutions we obtain in this way, cannot
satisfy, simultaneously, requirements set by energy conditions \emph{and}
localization of gravity on the brane. An exception to this result, is the case of
$\gamma=-1$ which gives a solution that can be translated to the one
arising within the scenario of \cite{rs2}. For this particular value
of $\gamma$, it is possible to have a regular solution (which is half of AdS$_5$) that trivially satisfies the null
energy condition and at the same time gives a finite four-dimensional Planck mass.

For a curved brane and a linear fluid, on the other hand, the situation
improves in the sense that, regular solutions now
become possible for a range of $\gamma$. Some of the regular solutions
can even satisfy the null energy condition; they correspond to AdS branes for $\gamma$ in the region $[-1,1/2)$. However, the problem of localizing
gravity on the brane met, previously, in the case of a flat brane, 
continues to arise and the only way to overcome it, is by compactifying
the bulk with a second brane as in \cite{rs1}.

Finally, for a flat brane and a non-linear equation of state, the situation is resolved:
we can construct a regular solution consistent with the null energy condition
which also localizes successfully gravity on the brane. It is an interesting open question whether such non-linear equation of state satisfying the null energy condition can be realised using an underlying microscopic description.

In this paper, we have also presented results which establish that there is a close connection between 3-brane setups in a five-dimensional bulk and cosmological solutions having a 4-dimensional spatial slice and evolving in proper time. This is accomplished by transforming the $Y\equiv Y_{OLD}$-dimension into  proper time, $Y_{OLD} \rightarrow it_{NEW}$, and the timelike dimension on the brane into a fifth spatial coordinate, $t_{OLD}\rightarrow -iY_{NEW}$, leaving the remaining three spatial coordinates intact. Therefore we can map from a braneworld, where everything depends on the transverse bulk space coordinate $ Y_{OLD}$, to a cosmological spacetime in $4+1$-dimensions evolving in time. This establishes a possible connection between cosmological phenomena and braneworld properties. In particular, the interpretation of the singular (in the transverse bulk coordinate) brane solutions, or those with some regularity (as discussed here), may correspond to cosmological solutions with special properties. For instance, the entropy of black holes in the braneworld may correspond to the cosmological entropy  of standard $(3+1)$- spacetime. 
We expect to analyse this point in a future publication.

\section*{Acknowledgments}
Work partially performed by I.A. as International professor of the Francqui Foundation, Belgium.

\end{document}